\documentstyle[preprint,aps]{revtex}
\begin{document}
\draft
\bibliographystyle{prsty}

\title{The path integral for Chern-Simons quantum mechanics}
\author{Silvio J. Rabello\thanks{e-mail: rabello@if.ufrj.br}\,
and Arvind N. Vaidya}

\address{\it Instituto de F\'\i sica, Universidade Federal do Rio de
Janeiro, Rio de Janeiro,  RJ \\ Caixa Postal 68.528-CEP 21945-970, Brasil}

\maketitle
\begin{abstract}

{\sl The path integral representation for a system of
N non-relativistic particles on the plane, interacting through a
Chern-Simons gauge field, is obtained from the operator formalism.
An effective interaction between the particles appears, generating  the
usual Aharonov-Bohm phases. The spin-statistics relation is also considered.}
\end{abstract}

The  unifying concept for planar systems with fractional statistics is the
long range, non-local mutual interaction of particles carrying an U(1)
charge  $e$ with ``statistical'' flux tubes at their location, in
an Aharonov-Bohm effect scenario \cite{frac}. A local field theoretical
description of these systems is provided if we couple the matter fields to a
gauge
field $A_\mu=({\bf A}({\bf x},t), A_0({\bf x,t}))$ with the dynamics given
by the topological Chern-Simons density \cite{Hagen}

\begin{equation}
\label{C-S}
{\cal L}(A)={\theta\over 2}\epsilon_{\mu\nu\lambda}A^\mu\partial^\nu
A^\lambda\,.
\end{equation}

This interaction has the effect of endowing the coupled particles with
flux tubes of strength $-{e\over\theta}$ , and the  statistics enters
through the Aharonov-Bohm phase that appears when one particle encircles
another, in addition to the intrinsic statistical one.

In this paper we obtain the path integral representation for the
transition amplitude of a  planar system with N non-relativistic particles,
in interaction with a Chern-Simons  gauge field. The effective action of
this path integral displays both the self-induced spin and
Aharonov-Bohm effects. To obtain this representation we perform the
canonical quantization of this system and after that apply a method
previously discussed in \cite{rafa} for  non-relativistic particles with
background gauge fields.

The Lagrangian for this planar system with N non-relativistic particles,
interacting with the Chern-Simons  gauge field is given by
\begin{equation}
\label{lagr}
L=\sum_{a=1}^N\biggl({m\over 2}{\dot {\bf x}}^2_a +e {\dot{\bf x}_a}\cdot
{\bf A }({\bf x}_a,t) -eA_0({\bf x}_a,t)
\biggr) +\theta\int d^2 x \biggl({1\over 2}\epsilon^{ij}{\dot A}_iA_j -
A_0B\biggr)\,.
\end{equation}
In order to quantize the above system we proceed  by
defining  the momenta for the point particles as
$p^i_a\equiv\partial L/ \partial{\dot x}_a^i$ and passing to the
Hamiltonian formalism. The corresponding quantum operators and
canonical commutation relations are given by  ($\hbar=1$)
\begin{eqnarray}
\label{co}
[{\hat A}_i({\bf x},t),{\hat A}_j({\bf y},t)]&=&
{i\over\theta}\epsilon_{ij}\delta({\bf x}-{\bf y})\,,\\
\label{xx}
[{\hat x}_i^a,{\hat x}_j^b]&=&0\,,\\
\label{xp}
[{\hat x}_i^a,{\hat \pi}_j^b]&=&i\delta^{ab}\delta_{ij}\,,\\
\label{Fij}
[ {\hat \pi}^a_i, {\hat \pi}^b_j]&=&
ie \epsilon_{ij}[{\hat  B}({\hat{\bf x}}_b)\delta_{ab}
+{e\over\theta}\delta ({\hat{\bf x}}_a-{\hat{\bf x}}_b)] \,,
\end{eqnarray}
with ${\hat \pi}_a^i\equiv {\hat p}_a^i-e{\hat A}^i(x_a)$
and ${\hat B}=\epsilon_{ij}\partial_i{\hat A}_j$. The ${\hat A}_0$
field has zero conjugate momentum being so  a Lagrange multiplier.
The time evolution is generated by the Hamiltonian operator
\begin{equation}
\label{Ham}
 {\hat H}=\sum_{a=1}^N \biggl({\hat{\mbox{\boldmath $ \pi$}_a}^2\over{2m}}
 +e{\hat A}_0(x_a,t)\biggr) +\theta\int d^2 x {\hat A}_0 (x,t){\hat B}(x,t)
 \,.
\end{equation}

Introducing the eigenvectors of ${\hat {\bf x}}_a(t)$
defined by ${\hat {\bf x}}_a(t)\vert x',t\rangle=x_a'\vert x',t\rangle$,
and the Schr\"{o}dinger wave functional $\Psi[A,t)$ for the
gauge fields we want to consider the transition amplitude
\begin{equation}
\label{prop}
G( {\bf x}'', {\bf x}';T)=\int [d\mu(A)]\Psi[A,0)^* \langle {\bf x}''
\vert e^{-i {\hat H}T}\vert {\bf x}'\rangle
\Psi[A,0)
\end{equation}
where $\vert x',0\rangle\equiv\vert x'\rangle$ and $[d\mu(A)]$ is
the integration measure over all the configurations of the gauge
field $A_\mu$. To obtain the wave functional $\Psi[A,t)$ we must
decide on which components of the gauge field  it depends since they
do not commute with one another.
We start with a convenient decomposition of the Schr\"{o}dinger
representation field
\footnote{ hereafter the fields with no time argument are evaluated at t=0}
${\hat{\bf A}}({\bf x})$  into longitudinal and transverse parts
\begin{equation}
\label{dec}
{\hat A}_i({\bf x})=\partial_i{\hat \xi}({\bf x})
-\epsilon_{ij}{\partial_j\over\nabla^2}{\hat B}({\bf x})\,.
\end{equation}
With this decomposition we have that (\ref{co}) leads to
\begin{equation}
\label{xb}
[{\hat \xi}({\bf x}),{\hat B}({\bf y})]=
-{i\over\theta}\delta({\bf x}-{\bf y})\,,
\end{equation}
Now we choose our wave functional to depend only on $\xi$
since the conjugate momentum to $A_0$ is zero. $\hat B$ acts on
it as the functional derivative $i\delta/\delta(\theta\xi)$.
The functional Schr\"{o}dinger equation for the gauge field coupled
to  N point-like sources as one can read from (\ref{Ham}) is \cite{dunne}
\begin{equation}
\label{sch}
i\partial_t \Psi[\xi ,t)=\int d^2 x \biggl[i\biggl(A_0
+{1\over\theta}J_i\epsilon_{ij}{\partial_j\over\nabla^2}\biggr)
{\delta\over\delta\xi}+ A_0\rho-\xi{\dot \rho} \biggr]\Psi[\xi,t)\,,
\end{equation}
with $J_i(x,t)=e\sum_{a=1}^N  {\dot x}_i^a(t)\delta({\bf x}-{\bf x}_a(t))$
and $\rho(x,t)=e\sum_{a=1}^N\delta({\bf x}-{\bf x}_a(t))$. The
solution is given by
\begin{equation}
\label{psi}
\Psi[\xi ,t)=exp\biggl[i\int d^2 x\biggl(\xi({\bf x})\rho({\bf x},t)
+\int^t d\tau J_i({\bf x},\tau)\epsilon_{ij}
{\partial_j\over\nabla^2}\rho({\bf x},\tau)\biggr)\biggl]
\end{equation}
As we can see  the above wave functional satisfies the Gauss law
\begin {equation}
\label{gauss}
\biggl[{\hat B}({\bf x})+{e\over\theta}\sum_{a=1}^N\delta({\bf x}
-{\bf x}_a(t))\biggr]\Psi[\xi,t) = 0\,,
\end{equation}
fixing the possible eigenvalues of the conjugate momentum of $\hat\xi$.
Since both ${\hat A}_0$ and $\hat\xi$ momenta are constrained we must,
in order to obtain a scalar product for $\Psi[\xi,t)$,  choose an
integration measure $[d\mu(\xi,A_0)]$ that selects one particular
configuration of the fields. Our choice is
$[d\xi][dA_0]\delta[\xi-\xi']\delta[A_0-A_0']$,
where $\xi'$ and $A_0'$ are arbitrary functions of x, that for the
sake of simplicity we set equal to zero everywhere.  With this wave
functional at hand we proceed to write a path integral for the
operator valued kernel $G( {\bf x}'', {\bf x}';T\vert {\hat{\bf A}})
\equiv\langle {\bf x}''\vert e^{-i {\hat H}T}\vert {\bf x}'\rangle$,
decomposing it  as  \cite {FeyHibbs}
\begin{equation}
\label{Dirac}
G( {\bf x}'', {\bf x}';T\vert {\hat{\bf A}})=\int d{\bf x}_p\dots
\int d{\bf x}_1 \langle {\bf x}'',T\vert {\bf x}_p,t_p\rangle
\langle{\bf x}_p,t_p\vert{\bf x}_{p-1},t_{p-1}
\rangle\dots \langle {\bf x}_1,t_1\vert {\bf x}',0\rangle\,.
\end{equation}
Then we take $T=(p+1)\varepsilon$ and $t_n=n\varepsilon$ $(n=1,\dots,p)$
letting $p\rightarrow\infty$ and $\varepsilon\rightarrow 0$ with T fixed.

To evaluate the small time propagator $\langle {\bf x}_{n+1},t_n
+\varepsilon\vert {\bf x}_n,t_n\rangle$
 we use a method developed by Schwinger
in his early investigations on effective actions \cite{Schw}. We write
the kernel $G( {\bf x}'', {\bf x}';T\vert {\hat{\bf A}})$ as  $exp(iW)$,
where $W({\bf x}'',{\bf x}';T\vert {\hat{\bf A}})$ is a complex functional
of ${\hat{\bf A}}$,
the end point coordinates and time. Defining the expectation value of an
observable $\hat{\cal O}$ by
\begin{equation}
\label{expect}
\langle {\hat{\cal O}}\rangle \equiv {\langle {\bf x}'',T\vert
{\hat{\cal O}}\vert {\bf x}',0\rangle /{\langle {\bf x}'',T\vert {\bf x}',
0\rangle }},
\end{equation}
it is easy to verify that $W$ is determined by the following equations
\begin{eqnarray}
\label{HJ}
\langle  {\hat H}( {\bf {\hat x}}(T), {\hat{\mbox{\boldmath $\pi$}}}(T))
\rangle &=&-{\partial W({\bf x}'',{\bf x}';T\vert {\hat{\bf A}})
\over{\partial T}} ,\\
\bigskip
\label{mom1}
\langle  {\hat \pi}^a_i(T)\rangle&=&{\partial W({\bf x}'',{\bf x}';
T\vert {\hat{\bf A}})\over{\partial  x''_{i,a}}}
+{e\over\theta}\epsilon_{ij}{\delta  W({\bf x}'',{\bf x}';T\vert
{\hat{\bf A}})\over\delta{\hat A}_j({\bf x}_a'')} -e {\hat A}_i({\bf x}_a'')
,\\
\bigskip
\label{mom2}
\langle  {\hat \pi}^a_i(0) \rangle&=&-{\partial W({\bf x}'',{\bf x}';T\vert
{\hat{\bf A}})\over{\partial x'_{i,a}}}-{e\over\theta}
\epsilon_{ij}{\delta  W({\bf x}'',{\bf x}';T\vert {\hat{\bf A}})
\over\delta{\hat A}_j({\bf x}_a')} -e {\hat A}_i({\bf x}_a') ,\\
\bigskip
\label{Norm}
W({\bf x}'',{\bf x}';0\vert {\hat{\bf A}})&=&-i\,
ln \delta^{2N} ({\bf x}''-{\bf x}').
\end{eqnarray}

To solve this problem Schwinger noticed that the above
equations relate the transition amplitude to the solution of the Heisenberg
equations for $ {\bf {\hat x}}_a (T)$ and
${\hat{\mbox{\boldmath $\pi$}}}_a(T)$.
If we solve for ${\hat{\mbox{\boldmath $\pi$}}}_a(T)$ in terms of
$ {\bf {\hat x}}_a(T)$ and $ {\bf{\hat x}}_a(0)$ and insert this, in a
time ordered fashion, in (\ref{HJ})-(\ref{mom2}) we are left with a set of
first order equations to integrate. For the small time kernel we must have
${\bf {\hat x}}_a(\varepsilon)$ up to second order in the small time
$\varepsilon$
\begin{equation}
\label{Heis}
{\hat x}^i_a(\varepsilon)=e^{i{\hat H}\varepsilon} {\hat x}^i_a (0)
e^{-i{\hat H}\varepsilon}\simeq {\hat x}^i_a(0)+{{\hat \pi}^i_a(0)\over m}
\varepsilon +{e\over 2m^2}\sum_{b=1}^N\epsilon^{ij}[{\hat B}({\hat{\bf x}}_b)
\delta_{ab}+{e\over\theta}\delta ({\hat{\bf x}}_a-{\hat{\bf x}}_b)
]{\hat \pi}^j_b(0)\varepsilon^2
\end{equation}
 Inverting the above equation
to get ${\hat \pi}^i_a(0)$ in terms of ${\bf {\hat x}}_a(\varepsilon)$ and
${\bf{\hat x}}_a(0)$
\begin{equation}
\label{pi0}
{\hat \pi}^i_a(0)\simeq m{\Delta{\hat  x}^i_a\over
\varepsilon}-{e\over 2}\sum_{b=1}^N\epsilon^{ij}[{\hat B}
({\hat{\bf x}}_b(\varepsilon) )\delta_{ab}+{e\over\theta}\delta
({\hat{\bf x}}_a (\varepsilon)-{\hat{\bf x}}_b(\varepsilon))]
\Delta{\hat  x}^j_b\,,
\end{equation}
where $\Delta{\hat  x}^i_a={\hat x}^i_a(\varepsilon)-{\hat x}^i_a(0)$.
Using the fact that $\langle (\Delta {\hat{\bf x}}_a)^2\rangle$ is of
order $\varepsilon$ \cite {FeyHibbs}, we see that if we take
$\langle{\hat{\mbox{\boldmath $\pi$}}}_a(0)\rangle$ the terms in the
above expansion are respectively of order $1/\sqrt{\varepsilon}$ and
$\sqrt{\varepsilon}$. The second  term although small in comparison with
the first, will give a relevant contribution when used in (\ref{mom1}) and
(\ref{mom2}) to evaluate $W({\bf x}_{n+1},{\bf x}_n;\varepsilon\vert
{\hat{\bf A}})$. From ${\hat{\mbox{\boldmath $\pi$}}}_a(0)$ we have by time
evolution
\begin{equation}
\label{pie}
{\hat \pi}^i_a(\varepsilon)\simeq m{\Delta{\hat  x}^i_a\over
\varepsilon}+{e\over 2}\sum_{b=1}^N\epsilon^{ij}
[{\hat B}({\hat{\bf x}}_b(\varepsilon) )\delta_{ab}+{e\over\theta}\delta
({\hat{\bf x}}_a(\varepsilon)-{\hat{\bf x}}_b(\varepsilon))]
\Delta{\hat  x}^j_b \,.
\end{equation}
Using the above expression for ${\hat {\mbox{\boldmath $\pi$}}}_a(0)$ or
${\hat {\mbox{\boldmath $\pi$}}}_a(\varepsilon)$ in ${\hat H}$,
in a time ordered manner, we are ready to integrate (\ref {HJ})
\begin{equation}
\label{W}
W({\bf x}_{n+1},{\bf x}_n;\varepsilon\vert {\hat{\bf A}})\simeq
\sum_{a=1}^N{m({\bf x}^{n+1}_a-{\bf x}^{n}_a)^2
\over 2\varepsilon} +iN \,ln\,\varepsilon +
\Phi({\bf x}_{n+1},{\bf x}_{n}\vert {\hat{\bf A}})\,,
\end{equation}
where we used that $[{\hat x}^i_a(\varepsilon),{\hat x}^j_b(0)]\simeq
-i\delta^{ij}\delta_{ab}\varepsilon/m$ and $\Phi$ is a time independent
functional of ${\bf A}$ and the end
point coordinates ${\bf x}^{n+1}_a$ and ${\bf x}^n_a$. Inserting the above
$W$ in (\ref{mom1}) and (\ref{mom2}) we have for $\Phi$ (remember
$\langle (\Delta {\hat{\bf x}}_a)^2\rangle\sim \varepsilon$)

\begin{eqnarray}
\label{phi}
{\partial\Phi \over{\partial x_{i,a}^{n+1}}}
+{e\over\theta}\epsilon_{ij}{\delta
\Phi\over\delta{\hat A}_j({\bf x}_a^{n+1})}
&=&{e\over 2}
\biggl [\Delta x_a^{k,n}{\partial {\hat A}_k({\bf x}^{n+1}_a)
\over{\partial x_{i,a}^{n+1}}}+ {\hat A}^i({\bf x}^{n+1}_a)
+{\hat A}^i({\bf x}^n_a)\\
&+& {e\over\theta}\sum_{b=1}^N\epsilon^{ij}
\delta^2 ({\bf x}^{n+1}_a-{\bf x}^{n+1}_b)\Delta x_b^{j,n}\biggr ]
+O(\varepsilon)\,,\\
\label{phi'}
{\partial\Phi \over{\partial x_{i,a}^{n}}}
+{e\over\theta}\epsilon_{ij}{\delta
\Phi\over\delta{\hat A}_j({\bf x}_a^n)}
&=&{e\over 2}
\biggl [\Delta x_a^{k,n}{\partial {\hat A}_k({\bf x}^{n}_a)
\over{\partial x_{i,a}^{n}}}-
{\hat A}^i({\bf x}^{n+1}_a)-{\hat A}^i({\bf x}^n_a)\\&+&
{e\over\theta}\sum_{b=1}^N\epsilon^{ij}\delta^2 ({\bf x}^{n}_a
-{\bf x}^{n}_b)\Delta x_b^{j,n}\biggr ]
+O(\varepsilon)\,,
\end{eqnarray}
with $\Delta{\bf x}_a^{n}={\bf x}^{n+1}_a-{\bf x}^n_a\,$.
Since in the rhs of the above equations we have
$O(\sqrt{\varepsilon})$ terms we are allowed to discard the
$O(\varepsilon)$ ones and easily find
the solution for $\Phi$
\begin{equation}
\label{phi2}
\Phi({\bf x}_{n+1},{\bf x}_n
\vert {\hat{\bf A}})=e
\sum_{a=1}^N\Delta{\bf x}^a_{n}\cdot
{\hat{\bf A}}({\bar{\bf x}}^a) + C\,,
\end{equation}
where  ${\bar{\bf x}}^a={1\over 2}({\bf x}^{n+1}_a+{\bf x}^n_a)$
and $C$ is the constant $-iN\, ln(m/2\pi i)$ determined by (\ref{Norm}),
setting $\varepsilon\rightarrow -i0_+$ in $e^{iW}$
and using the Gaussian representation for the delta function.

Putting this all together we get the small time kernel
\begin{equation}
\label{Prop2}
G( {\bf x}_{n+1}, {\bf x}_n;\varepsilon\vert {\hat{\bf A}})\simeq
\biggl({m\over{2\pi i\varepsilon}}\biggr)^N
exp\biggl[ i\sum_{a=1}^N \biggl({m(\Delta {\bf x}^a_{n})^2\over
2\varepsilon}+ e\Delta {\bf x}^a_{n}\cdot {\hat{\bf A}}({\bar{\bf x}}^a)
\biggr)\biggr]\,.
\end{equation}
Inserting it in (\ref{Dirac}) we get  the functional integral representation
for  $G( {\bf x}'', {\bf x}';T\vert {\hat{\bf A}})$,
\begin{equation}
\label{Prop3}
G( {\bf x}'', {\bf x}';T\vert {\hat{\bf A}})=\int [d^N{\bf x}]
exp\biggl(i\sum_{a=1}^N {m\over 2}{\dot{\bf x}}_a^2\biggr)\, {\bf T}
exp\biggl(ie\sum_{a=1}^N\int_0^T dt\, {\dot {\bf x}}_a
\cdot {\hat{\bf A}}({\bf x}_a)\biggr)\,,
\end{equation}
we here use the (abusive) continuum language where by
$[d^N{\bf x}]$ we mean the infinite product of terms
$ ({m/{2\pi i\varepsilon}})^Nd^N{\bf x}_k $ and the integration runs
from ${\bf x}(0)={\bf x'}$ to ${\bf x}(T)={\bf x''}$.
$\, {\bf T}exp\int {\hat {\bf A}}$ denotes the time ordered exponential
that stems naturally from (\ref{Dirac}). To undo this time
ordering in order to insert (\ref{Prop3}) in (\ref{prop}) we use the
BCH formula
\begin{equation}
\label{bch}
exp(M_k)\,exp(M_{k-1})\dots exp(M_0)=exp(\sum_n M_n)\, exp({1\over 2}
\sum_{m>n}[M_m,M_n])\,,
\end{equation}
($[M_m,M_n]$=c-number) that gives,
\begin{equation}
\label{Pord2}
 {\bf T}exp\biggl(ie\sum_{a=1}^N\int_0^T dt\,{\dot {\bf x}}_a\cdot
 {\hat{\bf A}}({\bf x}_a)\biggr)
 =exp(i\omega)\,exp\biggl(ie\sum_{a=1}^N\int_0^T dt\,
 {\dot {\bf x}}_a\cdot {\hat{\bf A}}({\bf x}_a)\biggr) \,,
\end{equation}
with
\begin{equation}
\label{om}
\omega=-{e^2\over 2\theta}\sum_{a,b=1}^N \int_0^T
dt\,\int_0^t d\tau\, {\dot x}^i_a(t)\epsilon_{ij}{\dot x}^j_b(\tau)\,
\delta({\bf x}_a(t)-{\bf x}_b(\tau))\,.
\end{equation}

Using the polarization (\ref{dec}) we can now evaluate the expectation value
in (\ref{prop}),
\begin{eqnarray}
\label{Pord4}
\int &[d\mu(A)]&\Psi[\xi,0)^*{\bf T}exp\biggl(ie\sum_{a=1}^N\int_0^T
dt\,{\dot {\bf x}}_a\cdot {\hat{\bf A}}({\bf x}_a)\biggr)\Psi[\xi,0)
\nonumber\\&=& exp\, i\biggl(\omega+{e^2\over\theta}\sum_{a,b=1}^N
\int_0^T dt\, {\dot x}^i_a\epsilon_{ij}{\partial_a^j\over\nabla_a^2}
\delta({\bf x}_a(t)-{\bf x}_b(0))\biggr)\,.
\end{eqnarray}
The argument of the exponential in the last line of (\ref{Pord4}) is
the effective interaction between the N particles. Noticing that
\begin{equation}
\label{om2}
\omega={e^2\over\theta}\sum_{a,b=1}^N\int_0^T dt\,
{\dot x}^i_a\epsilon_{ij}{\partial_a^j\over\nabla_a^2}
\biggl(\delta({\bf x}_a(t)-{\bf x}_b(t))-\delta({\bf x}_a(t)
-{\bf x}_b(0))\biggr)\,,
\end{equation}
we have for  the  $a\neq b$ case the following interaction
\begin{eqnarray}
S_{int}&=&{e^2\over\theta}\sum_{\stackrel {a,b=1}{a\neq b}}^N\int_0^T dt\,
{\dot x}^i_a\epsilon_{ij}{\partial_a^j\over\nabla_a^2}\delta({\bf x}_a(t)-
{\bf x}_b(t))\nonumber\\&=&
{e^2\over 2\pi\theta}\sum_{\stackrel {a,b=1}{a\neq b}}^N\int_0^T dt\,
{\dot x}^i_a\epsilon_{ij}{(x_a-x_b)^j\over \vert{\bf x}_a-{\bf x}_b\vert^2}
\,.
\end{eqnarray}
That is just the ordinary long range Aharonov-Bohm interaction between the
N particles. For $a=b$ we have the self-interaction terms
\begin{eqnarray}
\label{auto}
S_{self}&=&\lim_{t\rightarrow t'}{e^2\over\theta}\sum_{a=1}^N\int_0^T dt\,
{\dot x}^i_a\epsilon_{ij}{\partial_a^j\over\nabla_a^2}\delta({\bf x}_a(t)-
{\bf x}_a(t'))\nonumber\\&=&-{e^2\over 4\pi\theta}\sum_{a=1}^N \int_0^T
dt\,{d\over dt}tan^{-1}\biggl( {{\dot  x}^a_2 (t)\over{\dot  x}^a_1 (t)}
\biggr)\,.
\end{eqnarray}
To arrive at the last line we used the fact that
$\epsilon_{ij}\partial_i\partial_j tan^{-1}({\bf x})=
2\pi\delta({\bf x})$ and the identity \cite{dunne}
\begin{equation}
\label{tang}
\lim_{t\rightarrow t'}{d\over dt'}tan^{-1}\biggl({x_2(t')-x_2(t)
\over x_1(t')-x_1(t)}\biggr)={1\over 2}{d\over dt}tan^{-1}
\biggl({{\dot x}_2(t)\over {\dot x}_1(t)}\biggr)\,.
\end{equation}
The quantity in the last line of (\ref{auto}) is proportional to the sum of
angles swept by  ${\dot{\bf x}}_a(t)$ along the paths. Take for example N=1,
for  a closed smooth path with no self-intersections  we have a phase factor
$exp \,iS_{self}=exp\,i(-{e^2\over 2\theta})=
exp\,i\,2\pi s$, with $s$ defined as the spin of the particle,  giving the
celebrated spin-statistics relation $\vert s\vert ={e^2\over 4\pi\theta}\,$.
We achieved this result with no regulator for the $\delta$ function, if we
had used a regulator that preserves the odd nature of
$\partial_i\delta({\bf x})$ (e.g. Gaussian) such that
$\partial_i\delta(0)=0$ \cite{dunne}, there would be no spin contribution
at all. This ultraviolet ambiguity is a shortfall of the calculations
involving the Chern-Simons term.  In a recent paper \cite{ESGA} this
question was investigated, by avoiding the Chern-Simons term, in a
Berry phase calculation and no spin-statistics relation was found on
the plane, for the fractional quantum Hall effect quasiparticles.

The authors are grateful to Carlos Farina and Patricio Gaete
for reading the manuscript and for many stimulating discussions. This work
was partially supported by the CNPq (Brazilian Research Council).


\bigskip



\begin{thebibliography}{99}


\bibitem{frac} J.M. Leinaas and J. Myrheim, {\it Nuovo Cimento} {\bf 37 B}
(1977) 1 ; E.C. Marino and J.A. Swieca, {\it Nucl. Phys.} {\bf B 170}
(1980) 175 ; F. Wilczeck, {\it Phys. Rev. Lett.} {\bf 48} (1982) 1144.

\bibitem{Hagen} C. Hagen, {\it Ann. Phys.} (N.Y.) {\bf 157} (1984) 342.

\bibitem{rafa} S.J. Rabello and C. Farina, {\it Phys. Rev.} {\bf A 51}
(1995) 2614.

\bibitem{dunne} G. Dunne, R. Jackiw and C. Trugenberger, {\it Ann. Phys.}
(N.Y.) {\bf 194} (1989) 197.

\bibitem{FeyHibbs} R.P. Feynman and A.R. Hibbs, {\it Quantum Mechanics and
Path Integrals} (McGraw-Hill, New York, 1965).

\bibitem{Schw}J. Schwinger, {\it Phys. Rev.} {\bf 82}  (1951) 664.

\bibitem{ESGA} T. Einarsson, S.L. Sondhi, S.M. Girvin and D.P. Arovas,
{\it Fractional Spin for Quantum Hall Effect Quasiparticles},
preprint cond-mat/9411078.

\end{thebibliography}
\end{document}